\documentclass[a4paper,11pt]{article}
\pdfoutput=1 % if your are submitting a pdflatex (i.e. if you have
\usepackage{jcappub} % for details on the use of the package, please
\usepackage[T1]{fontenc} % if needed
\usepackage{multirow}

\title{\boldmath Viability of general relativity and modified gravity cosmologies using high-redshift cosmic probes}

\author[a,1]{Fernanda Oliveira\note{Corresponding author.}}
\author[a]{Bruno Ribeiro}
\author[b,c]{Wiliam S. Hip\'olito-Ricaldi}
\author[a]{Felipe Avila}
\author[a]{Armando Bernui}

\affiliation[a]{Observat\'orio Nacional, Rua General Jos\'e Cristino 77, 
S\~ao Crist\'ov\~ao, 20921-400 Rio de Janeiro, RJ, Brazil}
\affiliation[b]{Departamento de Ciências Naturais, CEUNES, Universidade Federal do Espírito Santo, Rodovia BR 101 Norte, km. 60, CEP 29.932-540, São Mateus, ES, Brazil}
\affiliation[c]{Núcleo Cosmo-UFES, CCE, Universidade Federal do Espírito Santo, Av. Fernando Ferrari, 540, CEP 29.075-910, Vitória, ES, Brazil}

\emailAdd{fernandaoliveira@on.br}
\emailAdd{brunoribeiro@on.br}
\emailAdd{wiliam.ricaldi@ufes.br}
\emailAdd{felipeavila@on.br}
\emailAdd{bernui@on.br}

\abstract{Several models based on General Relativity and Modified Gravity aim to reproduce the observed universe with precision comparable to the flat-$\Lambda$CDM cosmological model. In this study, we investigate the consistency of some of these models with current high-redshift cosmic data, assessing their ability to simultaneously describe both the background expansion and matter clustering, using measurements of the Hubble parameter $H(z)$, the luminosity distance $D_L(z)$, and the growth rate of structures $[f\sigma_8](z)$ through parametric and non-parametric methods. 
Our results indicate that background observables alone offer limited capacity to distinguish between models, while the inclusion of growth of structures data proves useful in revealing deviations, even if small. 
An $F(Q)$ model, the non-flat $\Lambda$CDM and  the $\omega$CDM emerge as alternatives well supported by data, closely matching the growth data and showing performance comparable to $\Lambda$CDM, as revealed by the Akaike Information Criterion. In contrast, $F(R)$ models are strongly disfavored compared to $\Lambda$CDM and $F(Q)$.  However, according to the Bayesian Information Criterion,  $\Lambda$CDM remains the preferred model among the models analysed.
These analyses illustrate the usefulness of both parametric and non-parametric approaches to explore the observational viability of alternative 
cosmological models.}

\begin{document}
\maketitle
\flushbottom

\section{Introduction}
\label{sec1:introducao}

The flat-$\Lambda$CDM, often termed the cosmological concordance model, successfully reproduces the features observed in diverse cosmological observables across different epochs of the universe's evolution; from primordial times, through the growth of matter clustering at large-scales, to the recent phase of accelerated expansion~\cite{Frieman2008,Basilakos2009,Weinberg2013}. 
However, the flat-$\Lambda$CDM model\footnote{In what follows we call this model simply $\Lambda$CDM.} is not the final model. There are still several issues to be elucidated, such as the physical nature of dark matter and dark energy, or the cosmological tensions, for instance~\cite{DiValentino2021,Perivolaropoulos2022, DiValentino2021s8,Adil2023,DiValentino2025}.

In recent years, several cosmological models have been proposed with the aim of addressing these problems, specially the one related to the {\it dark energy}, a mysterious fluid with an intriguing equation of state, fluid represented in the Friedmann equations by the cosmological constant $\Lambda$. 
The unknown physical nature of the dark energy is an uncomfortable situation that motivates the search for solutions within physical possible scenarios. The most promising scenario is to consider that General Relativity (GR) is not the final theory of gravitation, giving rise to the hypothesis that modified gravity (MG) theories could explain the recent phase of accelerated expansion of the universe~\cite{Capozziello2002, Tsujikawa2008, Clifton2011}.

On the one hand, these proposals include extensions of the $\Lambda$CDM model, as well as dynamical dark energy models based on GR. In fact, recent results suggest that GR-based models with an evolving equation of state constitute competitive alternatives to the $\Lambda$CDM model~\citep{DESI2025}. On the other hand, there are MG theories, such as $F(R)$~\citep{Clifton2011} and $F(Q)$~\citep{BeltranJimenez2019}, that are currently being tested~\footnote{Note that other approaches have also been explored, including for example, unified dark sector models, quintessence, dark matter-dark energy interacting models, other alternative formulations of gravity theories, etc. (see, e.g., \cite{Wang2024,Vargas2012,Funo2014,Oks2021,Tsujikawa2013,Cai2015,Bertini2019,Hipolito-Ricaldi2025}).}. Among the set of reported MG theories, generically termed by $F(R)$ or $F(Q)$, we decided to investigate the subset of viable theories, that is, those arising from a suitable modification of GR theory. This is because at Solar System scale, and also in the distant past $z \gg 1$, GR has passed with honors a set of astrophysical observations confirming its validity~\cite{Amendola2007,DeFelice2010,Ribeiro23}.

The so-called $F(R)$ theories are extensions of Einstein's GR theory, obtained by replacing the term $R - 2 \Lambda$ in the Einstein-Hilbert (EH) Lagrangian with a generic function of the Ricci scalar $R$, denoted $F(R)$. The $F(R)$ models offer alternative scenarios where the recent cosmic acceleration is an effect of the space-time geometry, making unnecessary the presence of the unknown, and exotic, {\it dark energy}. 
However, most of them were discarded for theoretical reasons, surviving those who reduce to GR in some limit, for this called viable models~\cite{Amendola2007}. 
The first successful $F(R)$ model was proposed by Starobinsky to analyze the inflationary era, proposing a model with the generic form $R + \epsilon R^2, \,\epsilon > 0$~\cite{Starobinsky1980}. 
Since then, several $F(R)$ models have been proposed, considering from simple polynomial laws to more complicated functions of the Ricci scalar (see, e.g., ~\cite{Amendola2007, Hu2007, Starobinsky2007, Appleby2007, Appleby2009, Tsujikawa2008, Cognola2008, Linder2009, Elizalde2011, Nautiyal2018, Oikonomou2013, Oikonomou2021}).

%Among the viable $F(R)$ models studied in the literature, we decided to investigate: Appleby-Battye, Hu-Sawicki, and Starobinsky. } 

In contrast, $F(Q)$ theories are generalizations of the symmetric teleparallel formulation of GR, which is characterized by a connection in which both curvature and torsion vanish, where gravity, instead, is described by a non-metricity scalar $Q$~\citep{Khyllep2021, Paliathanasis2025, BeltranJimenez2019, Lazkoz2019, Sahlu2024, Sahlu2025,Capozziello2022, DAgostino2022}. The $F(Q)$ model considered in this work is an alternative cosmological scenario with the interesting property that preserves the $\Lambda$CDM background spacetime, but modifies it at the perturbative level through the parameter $M$, which appears in the evolution equation of the matter fluctuations. 
It is a viable model that can be studied to describe observational data for both the recent phase of accelerated expansion of the universe and the clumpy universe~\citep{BeltranJimenez2019}.
%density contrast differential

Our aim in this work is to analyze the viability of alternative cosmological models and to search for observational hints that might shed light on the nature of dark energy or reveal a preference for GR or its possible modifications. We consider GR-based models and MG theories such as $F(R)$ and $F(Q)$, using cosmological data including cosmic chronometers (CC), Type Ia Supernovae (SNIa) at the background level, and  $[f\sigma_8](z)$, the normalized growth rate at the perturbative level. Our approach employs two well-known methods for statistical analysis: the Markov Chain Monte Carlo (MCMC) and Gaussian Processes (GPs)~\cite{Seikel2012, Jesus2019}. 

We use the cosmological data and MCMC analysis to obtain the best-fit parameter values for a given cosmological model. 
These results are then compared with model-independent reconstructions from GPs, allowing us to test the consistency and robustness of the MCMC fits without assuming any specific model. In fact, this comparison can reveal discrepancies suggestive of tensions between the model and the data. If the MCMC best-fit model falls to reproduce the GPs reconstructions, it may suggest that the model lacks the flexibility to capture all features in the data, or equivalently, that the data contain information  not explained by the model. After gaining some insight with this comparison, we use MCMC results along with the Akaike Information Criterion (AIC) and the Bayesian Information Criterion (BIC) to contrast different models based on GR and MG theories.

The models to be contrasted include the GR-based models: $\omega$CDM~\citep{Edesio2010}, with a constant equation of state for dark energy; the Chevallier-Polarski-Linder (CPL) parametrization, with a time-varying equation of state, $\omega_0 \omega_a$CDM~\citep{Chevallier2000,Linder2002}; and an extended $\Lambda$CDM including spatial curvature, $\Omega_k$-CDM.  We also consider viable $F(R)$ modified gravity models such as Starobinsky~\citep{Starobinsky2007}, Hu-Sawicki~\citep{Hu2007}, and the $R^2$-Corrected Appleby-Battye~\citep{Appleby2007, Appleby2009}. In addition, we study one $F(Q)$ model, that have a background evolution identical to GR~\citep{BeltranJimenez2019}. All these models are compared to the $\Lambda$CDM and rely on Friedmann-Lema\^{\i}tre-Robertson-Walker (FLRW) symmetries~\footnote{Although possible deviations from the cosmological principle have been reported~\citep{Kester2024,BOP2014}.}. From the observational point of view, several recent studies have constrained 
the parameters of these alternative models using cosmological data. 
In fact, GR-based models have been analyzed, for example, in ~\cite{DESI2025, Euclid2021, Afroz2025,  Xie2025, Planck2018, Luongo2024, DES2024}; 
$F(R)$ modified gravity models in~\cite{Nunes2017, Martinelli2009, Cardona2021, Kumar2025, Kumar2023, BHu2016, Bessa2021, Ribeiro23, Basilakos2013,Basilakos2017,Plaza2025}; and $F(Q)$ gravity in~\cite{Barros2020}. 
However, some of these studies focus exclusively on  one class of models (either GR, $F(R)$, or $F(Q)$), 
sometimes analyzing only the background data. 
Even when perturbative analyses are included, it is not common to find works that encompass all three types of gravities in a unified comparative framework. 
In this context, our contribution is to provide a consistent and up-to-date analysis by comparing GR, $F(R)$, and $F(Q)$-based models together, assessing both the background and growth of the structure analyses using the most recent available observational data. This allows for a more comprehensive evaluation of the viability of these competing gravitational scenarios. 

Recent observational studies have provided tight constraints on the parameters of the $F(R)$ and $F(Q)$ models considered here. For the Hu–Sawicki $(n=1)$ model, for example, using Baryon Acoustic Oscillations (BAO), Big-Bang Nucleosynthesis (BBN), and SNIa, Kumar et al.~\cite{Kumar2023} found \(b = -0.36 \pm 0.21\). More recently, Kumar et al.~\cite{Kumar2025} combined CC, SNIa from Union3.0, DESI 2025 BAO, Gamma-Ray Bursts (GRB), and the 
shift parameter and acoustic scale of the Cosmic Microwave Background (CMB), obtaining \(b = 0.217^{+0.084}_{-0.080}\). Plaza et al.~\cite{Plaza2025} used CC+SNIa+DESI2 BAO and found \(b = 0.000066^{+0.111}_{-0.000050}\). Independent constraints using Planck CMB + BOSS galaxy clustering + CMB lensing gave \(\log_{10}|f_{R0}| \lesssim -4\) \cite{Kou2024}, while cluster abundances from eRASS1 combined with weak-lensing (WL) mass calibration (DES, KiDS, HSC) yielded \(\log_{10}|f_{R0}| < -4.31\) (massless neutrinos) and \(\log_{10}|f_{R0}| < -4.12\) (massive neutrinos) at 95\% CL~\cite{Artis2024}.

We also highlight the forecasted constraints reported in the literature. For the Hu–Sawicki model with \(n=1\), Liu et al.~\cite{Liu2021} reported \(|f_{R0}| \lesssim 10^{-6}\), while Vogt et al.~\cite{Vogt2024} found \(\log_{10}|f_{R0}| < -5.95\) (SPT-3G) and \(\log_{10}|f_{R0}| < -6.23\) (CMB-S4) at 95\% CL. 
Davies et al.~\cite{Davies2024}, using weak-lensing peak statistics from FORGE and BRIDGE simulated surveys, showed sensitivity to \(\log_{10}|f_{R0}| \sim -6\) with \(\sim 2\%\) precision for Stage-IV surveys. 
Finally, the Euclid Collaboration~\cite{Euclid2025}, combining weak-lensing and photometric galaxy clustering, forecasted a sensitivity of \(\log_{10}|f_{R0}| \sim -5.6\) at 95\% CL.

For the Starobinsky model, Kumar et al.~\cite{Kumar2023} obtained \(b = 0.01 \pm 0.31\), Kumar et al.~\cite{Kumar2025} found \(b = 0.217^{+0.084}_{-0.080}\), and Plaza et al.~\cite{Plaza2025} reported \(b = 0.827^{+0.086}_{-0.119}\). 
Using \(H(z)\) and $[f\sigma_8](z)$ data, Bessa et al.~\cite{Bessa2021} obtained \(\lambda^{-1} = 1.460^{+0.47}_{-0.52}\) for the \(n=1\) model, and \(\lambda^{-1} = 0.99^{+0.72}_{-0.65}\) for the \(n=2\) model. 
Geng et al.~\cite{Geng2021}, considering a non-flat background with SNIa, CMB (Planck) and BAO, reported \(\lambda^{-1} < 0.283\) (68\% confidence level, CL) and \(\Omega_K = -0.00099^{+0.0044}_{-0.0042}\) (95\% CL). 
Chen et al.~\cite{Chen2019} obtained \(\lambda^{-1} = 0.282^{+0.099}_{-0.216}\) from CMB (Planck 2015), BAO, and SNIa from Supernova Legacy Survey (SNLS). 
For the \(R^2\)-corrected Appleby–Battye model, Ribeiro et al.~\cite{Ribeiro23} used CC+$[f\sigma_8](z)$+SNIa obtaining \(b = 2.18^{+5.41}_{-0.55}\).

For the $F(Q)$ model, Atayde and Frusciante~\cite{Atayde2023} constrained the total neutrino mass ($\sum m_\nu$) and the effective number of neutrino species ($N_{eff}$) using CMB, BAO, Redshift-Space Distortions (RSD), SNIa, galaxy clustering (GC), and WL. Using only CMB data, they found $\sum m_\nu < 0.731$ and $N_{eff} = 3.03^{+0.42}_{-0.39}$. Combining CMB+BAO+RSD+SNIa, they obtained $\sum m_\nu < 0.277$ and $N_{eff} = 2.93^{+0.31}_{-0.34}$. With all datasets (CMB+BAO+RSD+SNIa+GC+WL), they found $\sum m_\nu < 0.293$ and $N_{eff} = 2.91^{+0.31}_{-0.33}$. Using the same datasets~\cite{Atayde2021}, they also constrained the parameter $\alpha$, finding $\alpha = -0.64^{+0.64}_{-0.60}$ for CMB data only; $\alpha = -0.56^{+0.58}_{-0.57}$ for CMB+BAO+RSD+SNIa; and $\alpha = -0.05^{+0.34}_{-0.36}$ for CMB+BAO+RSD+SNIa+GC+WL.

For convenience, in some models we used slightly different parametric definitions to quantify deviations from GR. However, these values are fully consistent with, and directly mappable to, the latest observational constraints presented above.

This work is organized as follows: 
In Section \ref{data}, we briefly present the data and cosmological observables used. 
The Section~\ref{methodology} outlines the basics of the MCMC and GPs methodologies. 
In Section~\ref{alternatives}, we summarize main properties of the  alternative models analyzed in this work. Our results are presented and discussed in  Section \ref{results}, and finally, our conclusions are presented in Section~\ref{conclusions}.

\section{Cosmological observables and data}
\label{data}
%%%%%%%%%%%%%%%%%%%%%%%%%%%%%%%

Throughout this work, we consider three datasets: $H(z)$ measurements from CC~\citep{Niu24}, luminosity distance (or distance modulus) measurements from SNIa~\citep{Scolnic2021}, and $[f\sigma_8](z)$ measurements. 
Table containing CC is provided in Appendix~\ref{appendixA}, while the data for $[f\sigma_8](z)$ and SNIa can be found in \cite{Skara2019} and \cite{Pantheon2022}, respectively.

For $H(z)$, we use 31 measurements obtained using the cosmic chronometers technique (CC), a method that does not assume any cosmological model~\cite{Stern2009}. The CC technique estimates the derivative of redshift with respect to cosmic time by measuring the age difference between two ensembles of passively evolving massive galaxies at slightly different redshifts~\citep{Jimenez2001}. 
This approach avoids systematic errors associated with absolute age estimates by focusing on their relative age differences. The Hubble parameter is related to the derivative of redshift with respect to time by 
\begin{equation}
\label{CC}
H(z) = - \frac{1}{(1+z)} \frac{dz}{dt} \simeq - \frac{1}{(1+z)} \frac{\Delta z}{\Delta t} \,.
\end{equation}

For SNIa data, we use the Pantheon+ sample, which includes 1701 light curves of 1550 spectroscopically observed events and spectra collected across multiple surveys~\cite{Pantheon2022}. These data are used to determine the distance modulus $\mu$, defined as the difference between apparent and absolute magnitudes for each SNIa, i.e.,
\begin{equation} 
\label{eq:mu}
\mu \equiv m-M_B=5\log_{10}\left[\frac{D_{L}\left(z\right)}{1\,{\rm Mpc}}\right]+25\,,
\end{equation}
%where $d_{L}\left(z\right)$ is the luminosity distance, expressed as,
\begin{equation}
\label{eq:Dl}
D_{L}\left(z\right)=c\left(1+z\right)\int_0^z\frac{d\tilde{z}}{H\left(\tilde{z}\right)}\,.
\end{equation}
In our analyses, the data consist of the apparent magnitudes $m$, while the absolute magnitude $M_B$ is treated as a free parameter. 

In addition, for $[f\sigma_8](z)$ we use 35 uncorrelated measurements compiled by~\citep{Skara2019}.
The evolution of $[f\sigma_8](z)$ is described using linear perturbation theory, which governs the growth  density fluctuations through the matter density contrast $\delta_m(\textbf{r}, a)$~\cite{Coles1996,Avila21,Marques20,Franco25b}, 
defined as
\begin{equation}
\delta_m(\textbf{r},a) \equiv \frac{\rho_m(\textbf{r},a)-\Bar{\rho}_m(a)}{\Bar{\rho}_m(a)},
\end{equation}
where $\rho_m(\textbf{r},a)$ is the matter density at position 
$\textbf{r}$ and scale factor $a(t)$ at cosmic time $t$,  and $\Bar{\rho}_m(a)$ is the background matter density at same epoch. 

At sub-horizon scales, the evolution of the matter fluctuations is governed by the following second-order differential equation~\citep{Ribeiro23}
\begin{equation}\label{eq:edo}
\ddot \delta_m(t) + 2 H(t) \dot \delta_m(t) - 4 \pi G_{\text{eff}}\bar{\rho}_m(t) \delta_m(t) = 0,
\end{equation}
where $H(t) \equiv \Dot{a}(t)/a(t)$ is the Hubble parameter, dots denote derivatives  with respect to time and $G_{\text{eff}}\equiv G_{N}Q(t)$. Here, $G_{N}$ is the Newtonian gravitational constant, and the  function $Q(t)$ in Eq.~(\ref{eq:edo}) encodes deviation from GR. The effective gravitational coupling $G_{\text{eff}}$ is defined in the sub-horizon limit of linear perturbation theory through the modified Poisson equation appropriate for each gravity model considered in this work. Under GR theory one has $Q(t)=1$ and $G_{\text{eff}}\equiv G_{N}$. 
The explicit expression of $G_{\text{eff}}$ for the modified gravity models investigated in these analyses is provided in section \ref{alternatives}.

To solve Eq.~(\ref{eq:edo}), one must specify the cosmological model under study. 
In this work, we solve this equation for eight models: $\Lambda$CDM, $\omega$CDM, the CPL $\omega_0 \omega_a$CDM parametrization, $\Omega_k$-CDM, Starobinsky, Hu-Sawicki, $R^2$-Corrected Appleby-Battye, and one $F(Q)$ model,   to obtain their $[f\sigma_8]$ solution, given by
\begin{equation}
    [f\sigma_8](a) = \frac{\sigma_{8,0}}{\bar{\delta}_m(1)} \left[\frac{d \hspace{0.01cm} \delta_m(a)}{d \hspace{0.01cm} \ln(a)} \right],
\end{equation}
where $\sigma_{8,0}$ is the variance of the matter fluctuations at the scale of 8 $h^{-1}$Mpc evaluated at $z=0$~\citep{Coles1996,Franco25a}.

\section{Methodology} \label{methodology}
%%%%%%%%%%%%%%%%%%%%%%%%%%%%%%%%%%%%%%%%%%

In cosmology, the MCMC technique is commonly used to explore the parameter space of a specific cosmological model and to perform model comparison. 
As a model-dependent method, MCMC requires assumptions about the functional form of the cosmological probe being tested and its dependence on cosmological parameters, with the likelihood function determined by the model under investigation. 
In contrast, GPs are model-independent tools that do not rely on a specific functional form or parametric dependence. Instead, GPs use covariance functions (also called kernels) to model relationships directly from the data, enabling smooth and continuous reconstructions from observations. 
Our study combines both methods: first, we use MCMC to fit a cosmological model to the data, make predictions, and determine the best-fit for the given model. These results are then compared with reconstructions generated by GPs. This comparison helps us to assess the coherence, robustness, or any significant discrepancies, which may indicate tensions between the fitted model and the data. If the MCMC-fitted model fails to reproduce the GPs reconstruction, it suggests that the model does not fully capture the information contained in the cosmological data.

%%%%%%%%%%%%%%%%%%%%%%%%%%%%%%%%%%%%%%%%%%%%
\subsection{Markov Chain Monte Carlo}
%%%%%%%%%%%%%%%%%%%%%%%%%%%%%%%%%%%%%%%%%%%%

MCMC is a statistical tool used to construct a sequence of data points, called a chain, in parameter space to evaluate the posterior probability of a model (see, e.g., \cite{Padilla2019,Gilks1995,Gelman2013}). It combines Markov chains, which describe a sequence of states with probabilities depending only on the previous state, and Monte Carlo sampling, which uses randomness to sample from probability distributions. As noted in the literature, MCMC has several applications~\citep{Ribeiro23, Bessa2021, Mokeddem2025}. In this work, we use it for both parameter estimation and model comparison. The number of data points required to obtain reliable estimates with MCMC is known to scale linearly with the number of parameters, making this method significantly more efficient than grid-based approaches as the dimensionality increases~\cite{Padilla2019}.

According to Bayes' theorem, the posterior depends on the likelihood $\mathcal{L}(D|\vartheta,\alpha)$ and the prior $p(\vartheta,\alpha)$ via
\begin{equation}
P(\vartheta|D,\alpha) \propto \mathcal{L}(D|\vartheta,\alpha)\, p(\vartheta,\alpha)  \,,
 \end{equation}
where $\vartheta$ represents the set of model parameters, $D$ denotes the dataset, and $\alpha$ encodes the prior information model. 
In our analyses we use a flat prior and the likelihood is computed through 
\begin{equation}
\mathcal{L} \propto \exp(- \chi^2(D,\vartheta)/2) \,,
\end{equation}
where the chi-square, $\chi^2$, is given by 
\begin{equation}
\chi^2 = \sum_{ij} \Delta E_{i}^T C_{ij}^{-1} \Delta E_{j}\,,
\end{equation}
with
\begin{equation}
\Delta E_{i} \equiv E_i (\vartheta|\alpha) - D_i \,.
\end{equation}
Here, $E_i(\vartheta|\alpha)$ represents the $i$-th expected value based on the model, while $C_{ij}$ is the covariance matrix that encodes the statistical and systematic uncertainties associated with the dataset $D_i$. 

We then perform a joint likelihood analysis (SNIa+CC+$f\sigma_8$) to impose tighter constraints on the models over the redshift interval of the datasets. 
Thus, the total $\chi^2$ is given by
\begin{equation}
\chi^2 = \chi_{\rm SNIa}^2 + \chi_{\rm CC}^2 + \chi_{f\sigma_8}^2 \,,
\end{equation}
which results in the total likelihood 
\begin{equation}
\label{likelihoodtotal}
\mathcal{L} = \mathcal{L}_{SNIa} \times \mathcal{L}_{CC} \times \mathcal{L}_{f\sigma_8}\,.
\end{equation}
We use this analysis to estimate the best-fit value for the cosmological parameters for each model and to obtain the maximum likelihood, used to derive the $\chi^2_{min}$ and, consequently, the AIC~\cite{Akaike1974} and  BIC~\cite{Schwarz1978}.

%%%%%%%%%%%%%%%%%%%%%%%%%%%%%%%%
\subsection{Gaussian Processes}
%%%%%%%%%%%%%%%%%%%%%%%%%%%%%%%%

GPs have become one of the main tools for reconstructing cosmological quantities in a non-parametric manner~\citep{Seikel2012,Seikel13,Yang15,Valente18}. This method allows us to extract  information and make predictions from existing data independently of any underlying cosmological model. GPs have been used to study the evolution of various cosmological functions, such as the non-constant dark energy equation of state, $w(z)$~\cite{Seikel2012,Zhang18}, the deceleration parameter, $q(z)$~\citep{Jesus2019,Mukherjee2020}, $[f\sigma_{8}](z)$~\citep{Perenon21,Avila22b, Calderon2023msm, LHuillier2019imn}, the homogeneity scale, $R_{H}(z)$~\citep{Avila22a}, the distance duality relation, $\eta(z)$~\citep{Mukherjee21}, and a possible time evolution of the growth index, i.e., $\gamma = \gamma(z)$~\citep{Yin19,Avila22b,Mu2023, Oliveira2023}, among other applications in modern cosmology~\cite{Dinda2023,RuizZapatero2022,escamilla2023modelindependent,Sun2021,OColgain2021,Kjerrgren2021,Renzi2020, Calderon2022cfj, Dinda2023xqx, Oliveira25, Avila2025}.

Here, we employ a supervised learning regression approach to reconstruct $H(z)$, $D_L(z)H_0/c$ and $[f\sigma_8](z)$ in a non-parametric manner. GPs are a generalization of Gaussian distributions that characterize the properties of functions~\citep{Rasmussen06}. They are fully defined by their mean  and covariance functions, $m(\textbf{x})$ and $k(\textbf{x},\textbf{x}')$ respectively,
\begin{eqnarray}
 m(\textbf{x}) &=& \mathbb{E}[f(\textbf{x})],\nonumber\\
 k(\textbf{x},\textbf{x}') &=& \mathbb{E}[(f(\textbf{x})-m(\textbf{x}))(f(\textbf{x}')-m(\textbf{x}'))].
\end{eqnarray}
where $\mathbb{E}$ is the expected value for these functions. Then we write GPs as
\begin{equation}
    f(\textbf{x}) \sim \mathcal{GP}(m(\textbf{x}), k(\textbf{x},\textbf{x}')).
\end{equation}

Although GPs are  independent of an underlying cosmological model, they require a specific kernel to reconstruct $f(x)$. Here, we use the standard kernel: the Radial Basis Function, also known as the squared exponential (SE), given by
\begin{equation}\label{eq:SE_kernel}
    k(x, x') = \sigma_{f}^{2}\exp\left[-\frac{(x-x')^{2}}{2l^{2}}\right] \,,
\end{equation}
where $\sigma_{f}$ and $l$ are hyperparameters that are optimized during the reconstruction.

\section{Alternative cosmological models} \label{alternatives}
%%%%%%%%%%%%%%%%%%%%%%%%%%%%%%%%%%%%%%%%%%%%%
It is important to describe the alternative models analyzed in this work. These models are diverse: four are based on GR, three belong to the class of $F(R)$ theories, and one is based on $F(Q)$ theory. All are compared against the $\Lambda$CDM model. We  briefly discuss each of them in the following subsections.
%%-----------------------------------------------------------------------

%%--------------------------------------------

\begin{figure*}
\centering
\includegraphics[scale=0.69]{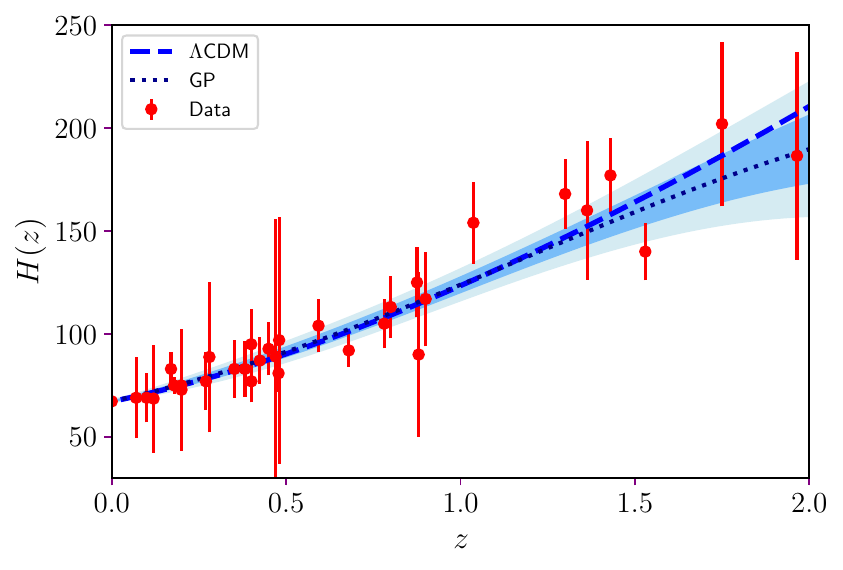} \,\,\,
\includegraphics[scale=0.69]{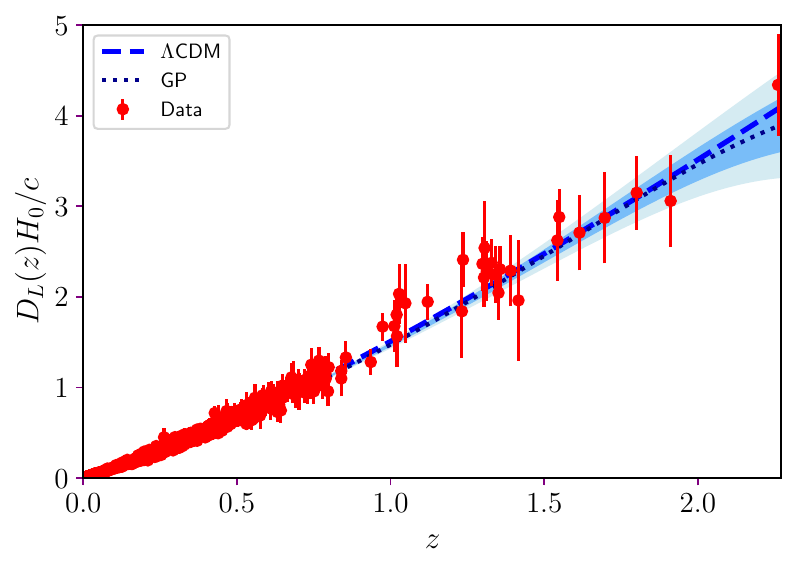} \,\,\,
\includegraphics[scale=0.69]{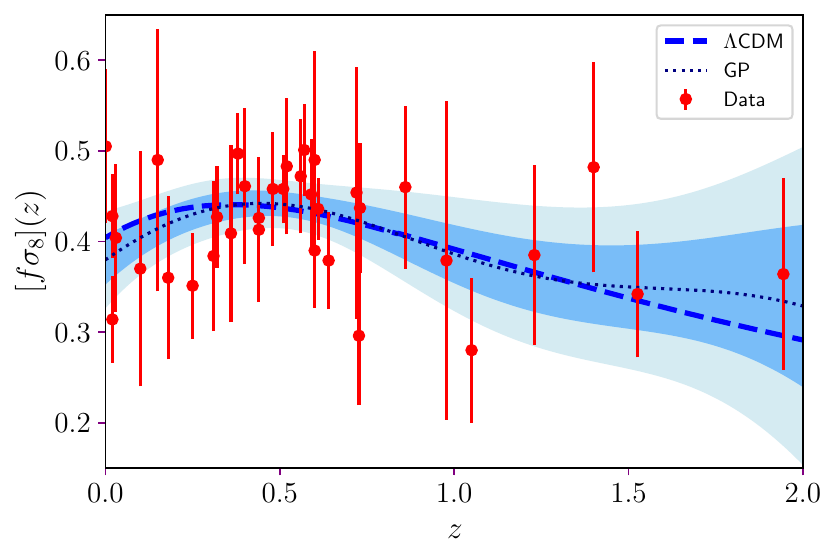}
\caption{
\textbf{Upper Panel:} $H(z)$ reconstruction from CC dataset. 
\textbf{Middle Panel:} $D_L(z)$ reconstruction from SNIa data. 
\textbf{Bottom Panel:} {$[f\sigma_8](z)$ reconstruction from the data}.
The shaded areas represent the 1$\sigma$ (dark blue) and 2$\sigma$ (light blue) CL regions.} 
\label{fig1}
\end{figure*}

%%-------------------------------
\subsection{GR-based cosmological models}
%$\Lambda$CDM-type cosmological models 

We examine models characterized  either by a different equation of state for dark energy or by  a non-Euclidean geometry of the universe. These models are based on GR, so the phase of accelerated expansion is still attributed to the presence of dark energy. 

In the $\omega$CDM model, the equation of state for dark energy, assumed to be $\omega = -1$ in $\Lambda$CDM, is still constant but treated as a free parameter, i.e., it is allowed to differ from $-1$. 
In this case, the Hubble parameter is 
\begin{equation}
\label{hubblewcdm}
H(a) = H_0 \sqrt{\Omega_{m0} a^{-3} + \Omega_{\Lambda 0} a^{-3(1+\omega)}} \,.
\end{equation}
%where $a$ is the scale factor in the Friedmann-Lema\^{\i}tre-Robertson-Walker metric.

For the CPL $\omega_0 \omega_a$CDM model, the dark energy equation of state is parametrized as
\begin{equation}
\omega(a) = \omega_0 + \omega_a (1-a) \,,
\end{equation}
where $\omega_0$ and $\omega_a$ are free parameters. 
In this case, the Hubble parameter is 
\begin{equation}
\label{hubblew0wacdm}
H(a) \,=\, H_0 \sqrt{\Omega_{m0} \,a^{-3}  \,+\, \Omega_{\Lambda 0} \,a^{-3(1+\omega_0 + \omega_a)} \,e^{-3 \omega_a (1-a)} } \,.
\end{equation}

The last GR-based model is $\Omega_k$-CDM. 
In this case, the geometry of the universe is no longer assumed to be 
Euclidean. 
Therefore, we have
\begin{equation}
\label{hubbleomk}
H(a) = 
H_0 \sqrt{\Omega_{m0}\,a^{-3} + \Omega_{k0}\,a^{-2} + \Omega_{\Lambda 0}} \,,
\end{equation}
where $\Omega_{m0}$, $\Omega_{k0}$, and $\Omega_{\Lambda 0}$ are the matter, curvature, and dark energy  density parameters, respectively.

%%-------------------------------
\subsection{F(R) theories}
$F(R)$ theories are probably the most investigated alternative models of gravity. These theories correspond to a modification of GR in which the term $R - 2\Lambda$ in the EH Lagrangian is replaced by a function of the Ricci scalar, $R$. Such a modification allows us to explain the current acceleration as an effect of the space-time geometry, without resorting to any exotic form of dark energy. On the other hand, an extra degree of freedom is introduced in the gravitational sector: the scalaron, a canonical scalar field whose potential is determined by $R$~\citep{DeFelice2010, Sotiriou2008, Nojiri2011, Nojiri2017}. The action for $F(R)$ gravity is written as~\citep{Capozziello2011, Clifton2011, DeFelice2010, Sotiriou2008, Ribeiro23}
\begin{equation}
\label{actionFR}
    S = \int d^4 x \sqrt{-g} \left[\frac{M_P ^2}{2}F(R) + \mathcal{L}_M \right],
\end{equation}
where $\mathcal{L}_M$ is the matter Lagrangian and $M_P ^2 \equiv (8 \pi G)^{-1}$ is the Planck mass.

In $F(R)$ gravity, $G_{\text{eff}}$ is modified by the presence of a scalar degree of freedom and, as shown in early works \cite{Tsujikawa2007, Pogosian2008,DeFelice2010,Bean2006}, can be expressed as
\begin{equation}
G_{\text{eff}} \equiv \frac{G}{F'(R)} \,\frac{1+ 4p (k^2/a^2R)}{1+3p (k^2/a^2R)}\,,
\end{equation}
where 
\begin{equation}
p \equiv \frac{R F''(R)}{F'(R)} \,,
\end{equation}
with $F'(R) \equiv dF(R)/dR$ and $k$ denotes the scale dependence. 
This expression follows from the scalar perturbation analysis in the subhorizon limit. To ensure compatibility  with the scale range probed by the $f\sigma_8$ data and to remain within the linear regime, we fix the scale to $k = $ 0.1 Mpc$^{-1}$ as also adopted in  \cite{Ribeiro23, Bean2006, Kumar2023, Basilakos2013, Tsujikawa2007}.

In this work, we examine several  $F(R)$ models. The first is the Starobinsky model, given by~\citep{Starobinsky2007}
\begin{equation}
\label{starobinski}
F_{\rm S}(R) \equiv R + \lambda R_s \left[ \left(1 + \frac{R^2}{R_s ^2} \right)^{-n} - 1 \right],
\end{equation}
where $R_s$, $\lambda$, and $n>0$ are the model parameters. In the high-curvature regime, $R \gg R_s$, the model yields an effective cosmological constant, $\Lambda \equiv \lambda R_s/2$. On the other hand, the present curvature scale $R_s$ is related to $\lambda$ via
\begin{equation}
R_s = \frac{6H_0^2 (1 - \Omega_{m0})}{\lambda}\,.
\end{equation}
The parameter $n$ is also related to $\lambda$, which leads to the following lower bounds: $\{n, \lambda_\text{min}\} = \{1, 1.54\}$, $\{2, 1.27\}$, $\{3, 1.04\}$, $\{4, 0.90\}$~\citep{Motohashi2010}. Usually the parameter $n$ is fixed. In this work, we consider $n=1$ and $n=2$, which are extensively studied cases in literature~\citep{Nunes2017,Kopp2013, Motohashi2009}, leaving $\lambda$ as the only free parameter. These choices ensure efficient chameleon screening and avoid tensions with Solar System tests, while also simplifying the parameter space by reducing degeneracies with other parameters~\citep{Bessa2021, Nunes2017}.

The second $F(R)$ model belongs to the commonly studied class of power-law theories, known as the Hu-Sawicki model, and is given by~\citep{Hu2007}
\begin{equation}
    F_{\rm HS}(R) \equiv R - m_0^2 \frac{c_1(R/m_0^2)^n}{c_2(R/m_0^2)^n + 1}\,,
\end{equation}
with $n>0$, and $c_1$ and $c_2$ being dimensionless parameters. The present curvature scale, $m_0$, is defined as
\begin{equation}
    m_0^2 \equiv \frac{8\pi G\rho_0}{3} = H_0^2\Omega_{m0}\,.
\end{equation}
The general relativistic limiting case, where $c_1/c_2^2 \rightarrow 0$ at fixed $c_1/c_2$, corresponds to an effective cosmological constant given by $\Lambda \equiv m_0^2 c_1/2c_2$. Since $\Lambda = 3H_0^2(1 - \Omega_{m0})$, we must have
\begin{equation}
c_1 = 6\,c_2\,\frac{1-\Omega_{m0}}{\Omega_{m0}} \,,
\end{equation}
linking the model parameters $c_1$ and $c_2$, which implies that the model has two free parameters: $n$ and $c_2$.

Alternatively, the  Hu-Sawicki model can be rewritten as
\begin{equation}
    \label{hu-sawicki}
    F_{\rm HS} (R) = R - 2\Lambda \frac{R^n}{R^n + \mu^{2n}}\,,
\end{equation}
by redefining
\begin{equation}
    \mu^2 \equiv m_0^2 c_2^{-1/n}\,.
\end{equation}
This form preserves the number of free parameters, now represented by $n$ and $\mu$. Note that since $\mu$ appears squared, only its absolute value $\left|\mu\right|$ is physically relevant. As in the Starobinsky case, $n$ is typically fixed in the Hu-Sawicki model, leaving $\mu$ as the only free parameter. Also similar to the Starobinsky case, such choices preserve chameleon efficiency and remain consistent with local gravity constraints~\citep{Bessa2021, Nunes2017}.

The last $F(R)$ theory considered here is the $R^2$-corrected Appleby-Battye ($R^2$-AB), given by~\citep{Appleby2009}
\begin{equation}
\label{appleby-batye}
F_{\rm AB}(R) \equiv \frac{R}{2} + \frac{\epsilon_{AB}}{2} \ln \left[\frac{\cosh(R/\epsilon_{AB} - b)}{\cosh b} \right] + \frac{R^2}{6 M^2},
\end{equation}
where $\epsilon_{AB}$ and $b$ are the model parameters, related via the equation
\begin{equation}
    \epsilon_{AB} = \frac{R_{vac}}{b + \ln\,(\,2\cosh b \,)}\,,
\end{equation}
where $R_{vac} \equiv 12H_0^2$ is the scalar curvature in vacuum. To reproduce the recent cosmic acceleration, the condition $b \geq 1.6$ is required~\citep{Appleby2009,Ribeiro23,Motohashi2012}. Finally, the parameter $M$ corresponds to the scalaron mass scale, which is determined from the amplitude of the primordial power spectrum as $M \approx 1.2 \times 10^{-5} M_P$ at inflation~\citep{Motohashi2012}.

%%-------------------------------
\subsection{F(Q) theories}
In these models the connection is not metric compatible anymore. The non-metricity tensor, defined as $Q_{\alpha \mu \nu} = \nabla_{\alpha} \,g_{\mu \nu}$, where $g_{\mu \nu}$ is the metric, is the fundamental object and gravity is described by the non-metricity scalar $Q$ \cite{Khyllep2021, Paliathanasis2025, BeltranJimenez2019, Lazkoz2019, Sahlu2024, Sahlu2025,Capozziello2022, DAgostino2022}
\begin{equation}
Q = - Q_{\alpha \mu \nu} P^{\alpha \mu \nu} \,,
\end{equation}
where $P^{\alpha \mu \nu}$ is the non-metricity conjugate defined as
\begin{equation}
P^{\alpha}_{\mu \nu} \equiv -\frac{1}{2} L^{\alpha}_{\mu \nu} + \frac{1}{4}(Q^\alpha - \tilde{Q}^\alpha) g_{\mu \nu} - \frac{1}{4} \delta^\alpha_{(\mu} Q_{\nu)} \,, 
\end{equation}
and $L^{\alpha}_{\mu \nu}$ is the disformation
\begin{equation}
L^{\alpha}_{\mu \nu} \equiv \frac{1}{2} Q^{\alpha}_{\mu \nu} - Q_{(\mu \nu)}^\alpha \,,
\end{equation}
where $Q_{\alpha}=g^{\mu\nu} Q_{\alpha\mu\nu}$ and $\tilde{Q}_{\alpha}=g^{\mu\nu}Q_{\mu\alpha\nu}$. 
Then, one can write the action for these theories as~\citep{BeltranJimenez2019, Khyllep2021}
\begin{equation}
\label{actionfQ}
S = \int d^4 x \sqrt{-g} \left[-\frac{1}{16 \pi G}F(Q) + \mathcal{L}_M \right] \,,
\end{equation}
where $\mathcal{L}_M$ is the matter Lagrangian. For the choice $F(Q) = Q$, one recovers GR.

In the case of $F(Q)$ gravity, the effective coupling $G_{\text{eff}}$ becomes \cite{Barros2020, BeltranJimenez2019}
\begin{equation}
    G_{\text{eff}} \equiv \frac{G}{F'(Q)}, 
\end{equation}
where $F'(Q) \equiv dF(Q)/dQ$
as derived in the metric-affine formulation of symmetric teleparallel gravity \cite{BeltranJimenez2019,Barros2020}.

In this work, we study one type of $F(Q)$ theory: one that reproduces GR's background evolution
but exhibits a modified evolution at the perturbative level. 
It is written as~\citep{Barros2020}
\begin{equation}
    \label{fQ1}
    F(Q) = Q + M\sqrt{Q} \,,
\end{equation}
where $M$ is given in units of $H_0$ and is related to a certain mass scale. This free parameter only appears at perturbative level, which means its value can only be determined with structure growth data.

\begin{figure*}
\centering
\includegraphics[scale=0.95]{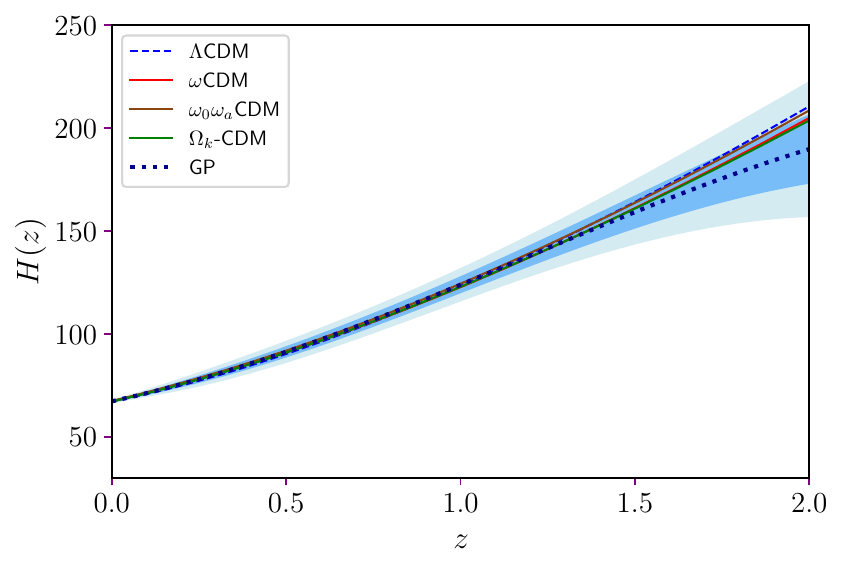} \,\,\,\,
\includegraphics[scale=0.95]{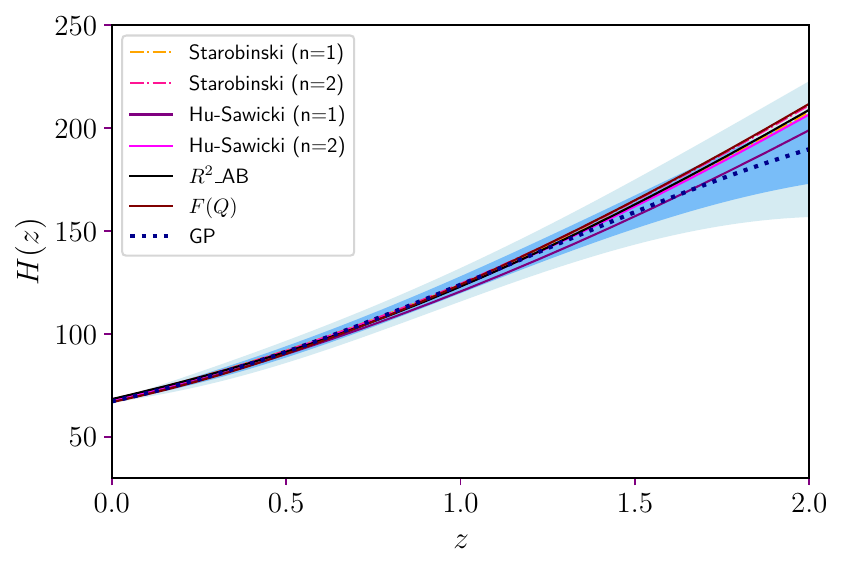}
\caption{
\textbf{Upper Panel:} Comparison among models based on GR for $H(z)$.
\textbf{Bottom Panel:} Same as in the left panel, but for MG models.
The shaded areas represent the 1$\sigma$ (dark blue) and 2$\sigma$ (light blue) CL regions.
See the text for a detailed discussion.} 
\label{fig2}
\end{figure*}

\begin{figure*}
\centering
\includegraphics[scale=0.95]{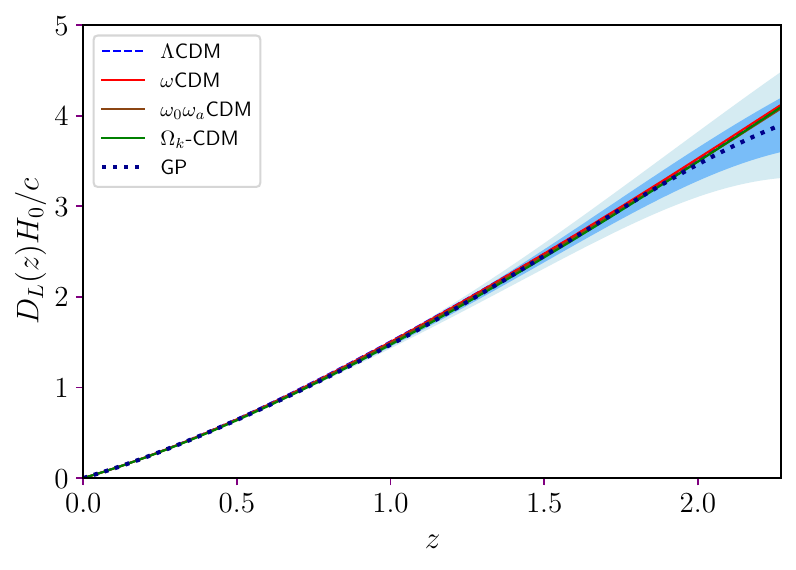} \,\,\,\,
\includegraphics[scale=0.95]{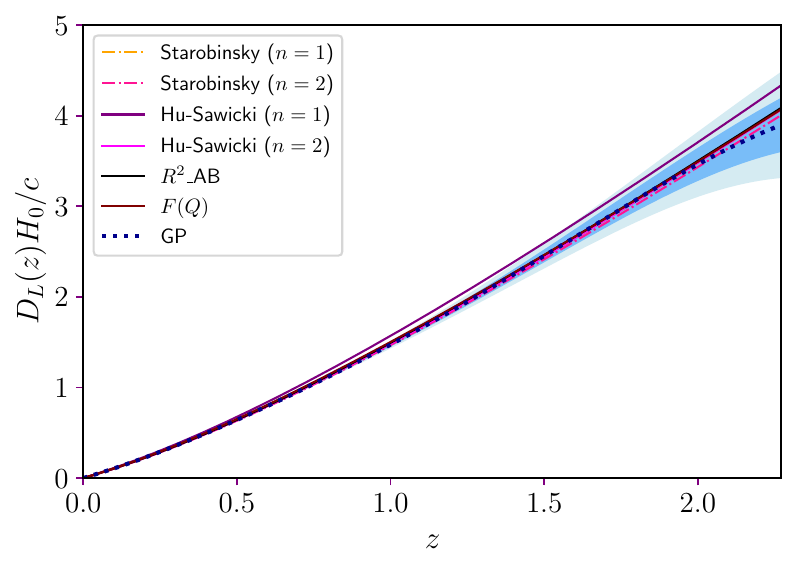}
\caption{
\textbf{Upper Panel:} Comparison among models based on GR for $D_L(z)H_0/c$.
\textbf{Bottom Panel:} Same as in the left panel, but for MG models.
The shaded areas represent the 1$\sigma$ (dark blue) and 2$\sigma$ (light blue) CL regions.
See the text for a detailed discussion.} 
\label{fig3}
\end{figure*}

\begin{figure*}
\centering
\includegraphics[scale=0.95]{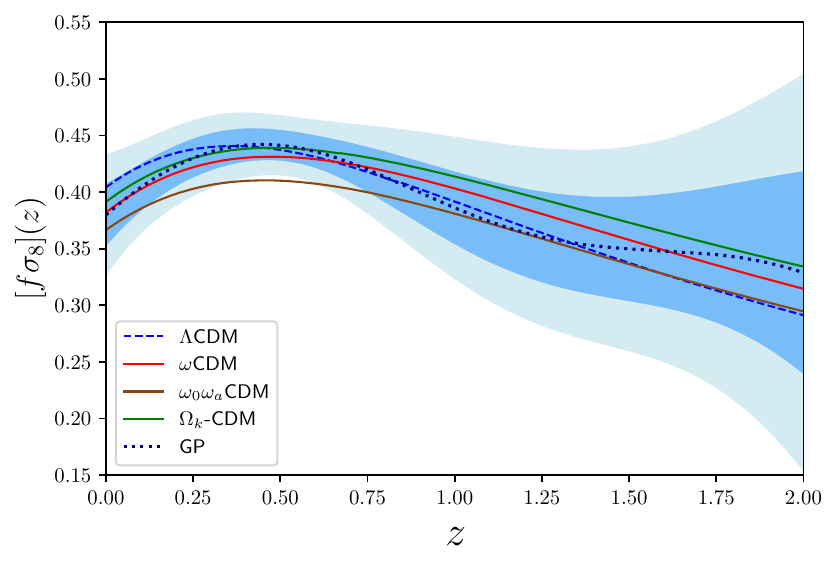} \,\,\,\,
\includegraphics[scale=0.95]{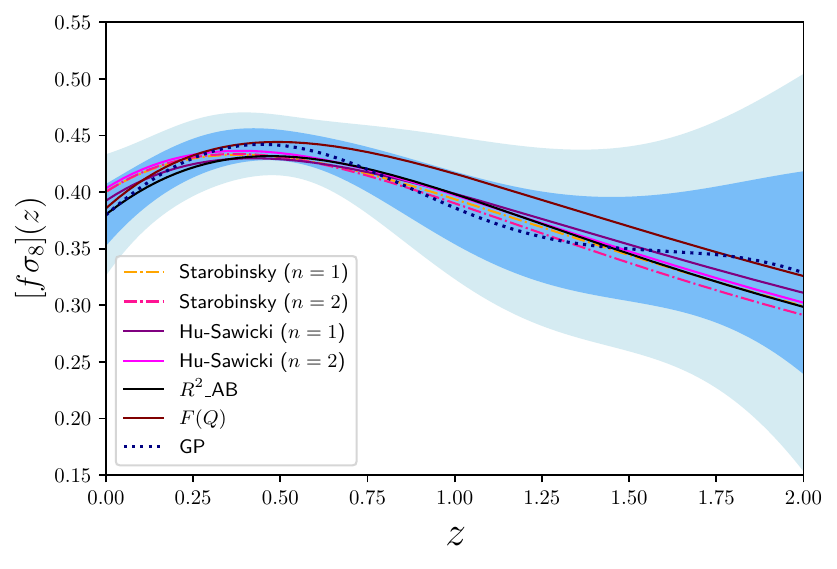}
\caption{
\textbf{Upper Panel:} Comparison among models based on GR for $[f\sigma_8](z)$.
\textbf{Bottom Panel:} Same as in the left panel, but for MG models.
The shaded areas represent the 1$\sigma$ (dark blue) and 2$\sigma$ (light blue) CL regions.
See the text for a detailed discussion.} 
\label{fig4}
\end{figure*}

\section{Results and Discussions}\label{results}

\begin{table}[tbp]
  \centering

  \resizebox{\columnwidth}{!}{
  
    \begin{tabular}{|l c c c c c c c c c|}

\hline
Model & $\,H_0$\, & $\,\Omega_{m0}\,$ & \,$\sigma_{8}$\, & $\,M_B\,$ & Model parameter & $\,\,\,\,\chi^2_{min}$ & $\,\,\,\,\chi^2_{\nu}$ &\,\,\,\, AIC & \,\,\, BIC \\
\hline
       $\Lambda$CDM  & $67.2^{+1.8}_{-1.7}$ & $0.34 \pm 0.02$ & $0.74 \pm 0.02$ & $-19.42 \pm 0.05$ & -- & 2913.42 & 1.65 & 2921.42 & 2943.33 \\ 
       
        $\omega$CDM  & $67.3 \pm 1.7$ & $0.31\pm 0.03$ & $0.77^{+0.05}_{-0.040}$ & $-19.41^{+0.06}_{-0.05}$ & $-0.89^{+0.06}_{-0.08}$ & 2912.76 & 1.65 & 2922.76 & 2950.14 \\
        
        $\omega_0 \omega_a$CDM & $67.1^{+2.0}_{-1.5}$ & $0.31\pm 0.03$ & $0.73^{+0.07}_{-0.03}$ & $-19.41^{+0.06}_{-0.05}$ & $-0.89\pm0.08$; $ -0.03^{+0.56}_{-0.61}$ & 2915.68 & 1.66 & 2927.68 & 2960.54 \\
        
       $\Omega_k$-CDM & $61.3^{+4.1}_{-3.1}$ & $0.32^{+0.03}_{-0.02}$ & $0.83\pm 0.06$ & $-19.42^{+0.06}_{-0.05}$ & $0.20^{+0.10}_{-0.13}$ & 2911.84 & 1.65 & 2921.84 & 2949.23\\
       
       Starobinsky ($n=1$) & $67.1 \pm 1.8$ & $0.33\pm 0.02$ & $0.75^{+0.03}_{-0.02}$ & $-19.41^{+0.06}_{-0.05}$ & $0.59^{+0.06}_{-0.31}$ & 2957.48 & 1.68 & 2967.48 & 2994.86 \\
       
       Starobinsky ($n=2$) & $66.4^{+1.9}_{-1.8}$ & $0.35\pm 0.02$ & $0.73\pm 0.02$ & $-19.42^{+0.05}_{-0.06}$ & $1.05^{+0.01}_{-0.63}$ & 3006.87 & 1.71 & 3016.87 & 3044.26 \\
       
       Hu-Sawicki ($n=1$) & $69.7^{+3.0}_{-2.7}$ & $0.25^{+0.06}_{-0.03}$ & $0.77^{+0.04}_{-0.03}$ & $-19.40\pm 0.05$ & $82^{+41}_{-44}$ & 2930.47 & 1.66 & 2940.47 & 2967.86 \\
       
       Hu-Sawicki ($n=2$) & $66.8^{+2.0}_{-2.3}$ & $0.33 \pm 0.02$ & $0.75 \pm 0.03$ & $-19.40^{+0.05}_{-0.06}$ & $104^{+45}_{-39}$ & 2918.18 & 1.66 & 2928.18 & 2955.57 \\
       
       $R^2$-AB & $67.3^{+2.0}_{-1.8}$ & $0.33 \pm 0.02$ & $0.74\pm 0.02$ & $-19.42^{+0.06}_{-0.05}$ & $1.98^{+5.03}_{-0.35}$ & 3022.65 & 1.72 & 3032.65 & 3060.03 \\
       
       $F(Q)$ & $66.9 \pm 1.8$ & $0.35 \pm 0.02$ & $0.80^{+0.04}_{-0.05}$ & $-19.42^{+0.05}_{-0.06}$ & $2.0^{+2.0}_{-1.4}$ & 2912.50 & 1.65 & 2922.50 & 2949.89\\
       %$F_2(Q)$ & 69.4 & ? & 0.703 & $-19.457$ & 1.608 & X & - & - \\
       \hline       
    \end{tabular}}
    \caption{Results of our MCMC analyses.}
    
    \label{tab:table1}
\end{table}

As a first step, we compute the key cosmological functions required for comparison with observational data: $H(z)$, $D_L(z)$, and $[f\sigma_8](z)$. 
For GR-based models, the Hubble parameter $H(z)$ admits an analytical expression (see equations~(\ref{hubblewcdm}),~(\ref{hubblew0wacdm}), and (\ref{hubbleomk})). On the other hand, for MG theories, $H(z)$ must be determined numerically. 
Once $H(z)$ is obtained, $D_L(z)$ is computed using equation~(\ref{eq:Dl}), 
and the matter density contrast is solved from equation~(\ref{eq:edo}) 
for each of the alternative models discussed in Section~\ref{alternatives}, with initial conditions set at $z=5$. Since the models must reproduce the matter-dominated era, which is necessary for structure formation, during early cosmic times ($z \gg 1$) they should not deviate significantly from the standard cosmological model.

Then, we perform a joint MCMC analysis using  three cosmological data sets, covering both the background and the clumpy universe: $H(z)$, $D_L(z)$, and $[f\sigma_8](z)$. The priors adopted were: 
$54 < H_0 < 76$, $0.1 < \Omega_{m0} < 0.5$, $0.6 < \sigma_{80} < 0.9$, and $-20.2 < M_B < -19$ for all models. 
For GR-based models, we used: $-5 < \omega < 0 $ for $\omega$CDM; $-1.5 < \omega_0 < 1.5$ and $-1.5 < \omega_a < 1.5$ for $\omega_0 \omega_a$CDM; and $-1 < \Omega_{k0} < 1$ for $\Omega_k$-CDM. 
For MG-based theories, the priors were: $0.0004 < \mu < 0.65$ for Starobinsky ($n=1$); $0.0001 < \mu < 1.06$ for Starobinsky ($n=2$); $0 < \mu < 300$ for Hu-Sawicki ($n=1$); $0 < \mu < 150$ for Hu-Sawicki ($n=2$); $1.6 < b < 12$ for $R^2$-AB; and $-3 < M < 3$ for $F(Q)$. 
Our MCMC analysis was implemented in Python using the~\textit{emcee} library, 
which is based on the {\em Affine Invariant Ensemble Sampler} 
algorithm~\cite{Goodman2010}. 
In all chains, we imposed the convergence criterion $R-1 < 10^{-2}$ for the Gelman-Rubin parameter $R$. Table \ref{tab:table1} summarizes our results, presenting the best-fit parameter values and their 1$\sigma$ uncertainties for all models discussed in Section~\ref{alternatives}, along with their corresponding minimum and reduced chi-square values, 
$\chi^2_{min}$ and $\chi^2_{\nu}$, respectively. As shown, the $\chi^2_\nu$ values are competitive across all models, exhibiting only slight variations among them.

To perform the GPs, we use the GaPP code\footnote{\url{https://github.com/JCGoran/GaPP}} , developed in \cite{Seikel2012}, which implements the algorithm from \cite{Rasmussen06}. Figure~\ref{fig1} shows the reconstructions of $H(z)$, $D_L(z)H_0/c$, and $[f\sigma_8](z)$ at the 1$\sigma$ and 2$\sigma$ confidence levels (CL) across the redshift ranges of the corresponding datasets. The dashed blue line represents the $\Lambda$CDM model. Note that in all cases, $\Lambda$CDM lies well within  the 1$\sigma$ region. The increase in uncertainties in regions with sparse data may  possibly cause the apparent deviations observed in those regions.

In Figure \ref{fig2}, we present the comparison of the models for $H(z)$, along with its GPs reconstruction at the 1$\sigma$ and 2$\sigma$ CL over the redshift range $z$ $\in$ [0.0, 2.0]. The curves correspond to the best-fit values listed in Table \ref{tab:table1}. The left panel displays the models based on GR, while the right panel shows the models based on MG. All models remain within the 1$\sigma$ CL up to $z \sim 1.5$ \footnote{For redshift values higher than 1.5, the sparse data may bias our comparisons.}. This indicates that, at current data precision, $H(z)$ alone is not sufficient to discriminate among the studied models. Figure~\ref{fig3} shows a similar comparison, but for $D_L(z)H_0/c$, along with its GPs reconstruction at the 1$\sigma$ and 2$\sigma$ CL over the redshift interval $z \in [0., 2.27]$. Again, the left panel corresponds to GR-based models, while the right panel shows the MG-based models. Almost all models remain within the 1$\sigma$ CL region, however, the Hu-Sawicki model ($n=1$) is an exception, as it lies  outside of the 1$\sigma$ region but still remains within of the 2$\sigma$ region.

Figure \ref{fig4} shows the comparison among models using the $[f\sigma_8](z)$ data and the GPs reconstruction along with its predictions at 1$\sigma$ and 2$\sigma$ CL over the redshift range $z$ $\in$ [0.0, 2.0]. The left panel displays the models based on GR, while the right panel presents the models based on MG. As can be seen, the MG-based models qualitatively perform slightly  better than the GR-based one, as they tend to follow the reconstructed $[f\sigma_8](z)$ curve more closely, remaining consistently within the 1$\sigma$ CL. On the other hand, the cosmological model with dynamic dark energy in the CPL parametrization, $\omega_0\omega_a$CDM, is the only model that slightly departs from the 2$\sigma$ CL in the interval $0.3 \lesssim z \lesssim 0.6$, while still remaining within the 1$\sigma$ - 2$\sigma$ CL in the rest of the redshift range in analysis.

%It is important to emphasize that although 
An illustrative comparison is presented in Figures~\ref{fig2}, \ref{fig3}, and \ref{fig4}, which allows us to check the consistency of MCMC fits without favoring any specific model. 
However, it is important to emphasize that these comparisons with GPs reconstructions do not account for the complexity of the models, which involve different numbers of free parameters. Naturally, increasing the number of free parameters can make it easier for a model to fit the data, potentially giving a misleading impression of agreement. 
Therefore, to ensure a fair and objective comparison among models, it is necessary to return to the MCMC results and apply statistical tools, such as the Akaike Information Criterion (AIC)~\citep{Akaike1974} and the Bayesian Information Criterion (BIC)~\citep{Schwarz1978}.

The $\chi^2$ statistic is suitable for identifying the best-fitting parameters within a model, but is not suitable for comparing models with different numbers of parameters, as lower $\chi^2$ values can be achieved by increasing the model complexity. A more robust comparison can be made using the Akaike Information Criterion~\citep{Akaike1974}, which penalizes models with more free parameters. The numerical AIC is computed as 
\begin{equation}
\text{AIC} \equiv \chi^2_{min} + 2k \,,
\end{equation}
where $k$ is the number of free parameters in the model. Thus, a lower AIC value generally means that the model fits the data better. However, a more important methodology for model comparison is to calculate the relative difference between models, i.e, $\Delta \text{AIC}$. We examined three cases, whose results are summarized in Table~\ref{AICcomparison}: 

\begin{itemize}
\item 
%most widely accepted
Comparison with $\Lambda$CDM: Since $\Lambda$CDM is the 
concordance cosmological model, we compare all the models discussed with it. 
Our results indicate that, based on the observables we used, three models are competitive with $\Lambda$CDM, namely,  $\Omega_k$-CDM, $F(Q)$ and  $\omega$CDM, while $F(R)$ models are strongly disfavored.
\item 
Comparison of  MG models with $F(Q)$:  
We compare all MG models with $F(Q)$, which appears as 
the  most competitive MG-based model relative to 
$\Lambda$CDM. Our results indicate that almost all the $F(R)$ models are strongly disfavored except the Hu-Sawicki ($n=2$) model, which is considerably less favored than $F(Q)$. 
\item 
Comparison between $F(R)$ models and the Hu-Sawicki model ($n=2$): We compare the $F(R)$ models with the Hu-Sawicki model ($n=2$), which has the lowest  AIC among the $F(R)$-based models. All the $F(R)$ are strongly disfavored.
\end{itemize}

\begin{table}[tbp]
    \centering

\label{AICcomparison}

\resizebox{\columnwidth}{!}{
\begin{tabular}{|l l c c|} 
\hline
\textbf{Comparison} & \textbf{Model} & \textbf{$\Delta$AIC} & \textbf{Empirical Support}  \\
\hline
\multirow{3}{*}{\textbf{$\Lambda$CDM vs. all models}} 
& $\Omega_k$-CDM &  0.42 &  Substantial  \\
& $F(Q)$ & 1.08 & Substantial  \\
& $\omega$CDM  & 1.34 & Substantial \\
&  $\omega_0 \omega_a$CDM   & 6.26 & Considerably Less  \\
&  Hu-Sawicki ($n=2$)   & 6.76 & Considerably Less \\
&  Hu-Sawicki ($n=1$)   & 19.05 &  Essentially None \\
&  Starobinsky ($n=1$)   & 46.06 &  Essentially None \\
&  Starobinsky ($n=2$)   & 95.45 &  Essentially None \\
&  $R^2$-AB   & 111.23 &  Essentially None \\
\hline

\multirow{3}{*}{\textbf{$F(Q)$ vs. $F(R)$ models}} 
& Hu-Sawicki ($n=2$)  & 5.68 & Considerably Less   \\
& Hu-Sawicki ($n=1$) & 17.97 & Essentially None \\
& Starobinsky ($n=1$) & 44.98 & Essentially None  \\
& Starobinsky ($n=2$) & 94.37 & Essentially None  \\
&  $R^2$-AB & 110.15 & Essentially None \\
\hline

\multirow{3}{*}{\textbf{HS\,($n=2$) vs. $F(R)$ models}} 
& Hu-Sawicki ($n=1$) & 12.29 & Essentially None  \\
& Starobinsky ($n=1$) & 39.3 & Essentially None \\
&  Starobinsky ($n=2$) & 88.69 & Essentially None\\
&  $R^2$-AB & 104.47 & Essentially None  \\
\hline
\end{tabular}} 
\caption{Statistical comparison among models using AIC, with qualitative interpretation according to the calibrated Jeffreys' scale \cite{Perez-Romero2017}}.

\label{tab:table2}

\end{table}

Another robust criterion for model comparison is the Bayesian Information
Criterion (BIC)~\citep{Schwarz1978},  which properly accounts for dataset size and penalizes model complexity more strongly than the AIC. BIC is computed as 
\begin{equation}
    \text{BIC} \equiv \chi^2_{min} + k \ln{n} \,,
\end{equation}
where $k$ is the number of free parameters in the model and 
$n$ is the sample size. As we did for the AIC, we calculated the relative differences between models, $\Delta$BIC, and examined the same three cases. The results are shown in Table \ref{tab:table3}. The conclusions based  on AIC change slightly when using BIC. Compared to $\Lambda$CDM, BIC indicates moderate (positive) evidence against the $\Omega_k$-CDM model ($\Delta BIC=5.9$), while $F(Q)$ and $\omega$CDM, which were competitive under AIC, are now strongly disfavored ($\Delta BIC = 6.56$ and $6.81$, respectively).  All remaining models are very strongly disfavored ($\Delta BIC >10$). When $F(R)$ models are compared with $F(Q)$, BIC presents moderate (positive) evidence against the Hu–Sawicki ($n=2$) model ($\Delta BIC = 5.68$), while all other $F(R)$ models are very strongly disfavored ($\Delta BIC >10$). Finally, when Hu–Sawicki ($n=2$) is compared with the remaining $F(R)$ models, BIC indicates very strong evidence against all of them ($\Delta BIC >10$).

\begin{table}[tbp]
    \centering

\label{BICcomparison}

\resizebox{\columnwidth}{!}{
\begin{tabular}{l l c c } 
\hline
\textbf{Comparison} & \textbf{Model} & \textbf{$\Delta$BIC} & \textbf{Evidence Against Model}  \\
\hline
\multirow{3}{*}{\textbf{$\Lambda$CDM vs. all models}} 
& $\Omega_k$-CDM &  5.9 &  Positive  \\
& $F(Q)$ & 6.56 &  Strong \\
& $\omega$CDM  & 6.81 & Strong \\
&  Hu-Sawicki ($n=2$)   & 12.24 & Very Strong \\
&  $\omega_0 \omega_a$CDM   & 17.21 &  Very Strong \\
&  Hu-Sawicki ($n=1$)   & 24.53 & Very Strong  \\
&  Starobinsky ($n=1$)   & 51.53 &  Very Strong \\
&  Starobinsky ($n=2$)   & 100.93 & Very Strong  \\
&  $R^2$-AB   & 116.7 &  Very Strong \\
\hline

\multirow{3}{*}{\textbf{$F(Q)$ vs. $F(R)$ models}} 
& Hu-Sawicki ($n=2$)  & 5.68 &  Positive  \\
& Hu-Sawicki ($n=1$) & 17.97 &  Very Strong \\
& Starobinsky ($n=1$) & 44.97 & Very Strong  \\
& Starobinsky ($n=2$) & 94.37 & Very Strong  \\
&  $R^2$-AB & 110.14 &  Very Strong \\
\hline

\multirow{3}{*}{\textbf{HS\,($n=2$) vs. $F(R)$ models}} 
& Hu-Sawicki ($n=1$) & 12.29 &  Very Strong  \\
& Starobinsky ($n=1$) & 39.29 & Very Strong  \\
&  Starobinsky ($n=2$) & 88.69 & Very Strong  \\
&  $R^2$-AB & 104.46 & Very Strong   \\
\hline
\end{tabular}}

\caption{Statistical comparison among models using BIC, with qualitative interpretation according to the calibrated Jeffreys' scale \cite{Perez-Romero2017}}.

\label{tab:table3}

\end{table}

%%-------------------------------------------------------------------
\section{Conclusions}\label{conclusions}

The comparison between cosmological data and predicted functions from theoretical models remains the main approach to testing alternatives to the $\Lambda$CDM paradigm. 
Although the concordance model has successfully described a wide range of cosmological observations, it still faces open challenges, 
like understanding and modeling the evolution of the cosmic growth of structures, 
from the primordial density fluctuations to the current observed universe~\citep{Coles1996,Basilakos2009,Skara2019,Franco24}.

In this work, we investigated several alternative models, including $\omega$CDM, $\omega_0 \omega_a$CDM, $\Omega_k$-CDM, Starobinsky, Hu-Sawicki, $R^2$-Corrected Appleby-Battye, and one $F(Q)$ model, using both GPs and MCMC approaches. 
We used data from $H(z)$, $D_L(z)$, and $[f\sigma_8](z)$, and applied the MCMC methodology to infer the  best-fit parameter values for each model, subsequently comparing them with GPs reconstructions to assess their consistency with the data. 
Our results confirm that background observables alone, namely $H(z)$ and $D_L(z)$, are in general insufficient to discriminate between models, as they show similar behavior within current uncertainties. 
Furthermore, one observes in Figure~\ref{fig3} that the $D_L(z)$ curve for Hu-Sawicki ($n=1$) model deviates at more than 1$\sigma$ confidence level from the GPs reconstruction using SNIa data, although this model remains consistent  at  2$\sigma$. On the other hand, the inclusion of cosmic growth data proved to be particularly useful, as it highlighted deviations of some models from the GPs reconstruction, as in the case of the CPL $\omega_0 \omega_a$CDM model which slightly departs from 2$\sigma$ (in a small part of the redshift range), as shown in Figure~\ref{fig4}.

Additionally, we performed a quantitative model comparison using the AIC and the BIC to account for an appropriate penalization of the complexity of the model. Our results for AIC, summarized in Table~\ref{tab:table2}, indicate that the $F(Q)$ model emerges as a competitive alternative, alongside two  GR-based models: $\Omega_k$-CDM and $\omega$CDM, while $F(R)$ theories are strongly disfavored when compared to the flat-$\Lambda$CDM and $F(Q)$ models. 
This is likely due to the fact that $F(Q)$ model preserves GR-like background dynamics while modifying the growth of structures in a way that remains consistent with current data. Furthermore, our comparisons highlighted that the $F(Q)$ model closely follows the GPs reconstructed $[f\sigma_8](z)$ function, reinforcing its viability as a potential alternative to the standard cosmological model. 
Moreover, considering data from the perturbed universe, the $\omega_0 \omega_a$CDM model appears to exhibit a mild tension with the GPs reconstructions, and when compared to the flat-$\Lambda$CDM model it receives considerably less support from the data. %, as observed by the application of the AIC. 
For the BIC, our results, summarized in Table~\ref{tab:table3}, confirm that, under a more conservative penalization for model complexity, $\Lambda$CDM remains the preferred model among those tested, and all alternatives show at least moderate (positive) evidence against them.

Overall, the recent constraints reported in the literature and presented in Section~\ref{sec1:introducao} indicate that deviations from GR, if they exist, must be very small. 
In this context, our results (see  Table \ref{tab:table1}) 
are in agreement with the latest observational bounds and, importantly, the statistical comparison done using information criteria (AIC/BIC) places the $F(R)$ models at a clear disadvantage with respect to GR (see Tables \ref{tab:table2} and \ref{tab:table3}), highlighting a key outcome of our study by 
directly comparing a variety of modified gravity theories.

Our final conclusion is that this study shows that both parametric and non-parametric methodologies offer complementary and valuable approaches to explore the observational viability of alternative cosmological models, providing a more comprehensive perspective on their performance in light of current data. 

\acknowledgments
FO and BR thank CAPES for their fellowships. WSHR acknowledges FAPES. FA thanks to Fundação de Amparo à Pesquisa do Estado do Rio de Janeiro (FAPERJ), Processo SEI-260003/001221/2025,  for the financial support.  
AB acknowledges a CNPq fellowship.

\appendix

%%---------------------------------------------
\section{Appendix: Data Sets}
\label{appendixA}

In this appendix, we present the CC dataset used in this work. Table~\ref{tab:tableA0} lists the 31 measurements of $H(z)$ obtained through the CC method, as compiled by~\cite{Niu24}. 

\begin{table} [htb!]
\centering
    \begin{tabular}{|l | l | l | l|}
\noalign{\smallskip}\hline\noalign{\smallskip}
$z$       & $H(z)$ [km/s/Mpc]          &  $z$     &     $H(z)$ [km/s/Mpc]          \\
        \hline
        $0.07$    &  $69.0\pm 19.6$  &  $0.4783$  &  $80.9\pm 9.0$    \\   
        $0.09$    &  $69.0\pm 12.0$  &  $0.48$    &  $97.0\pm 62.0$   \\     
        $0.12$    &  $68.6\pm 26.2$  &  $0.593$   &  $104.0\pm 13.0$  \\  
        $0.17$    &  $83.0\pm 8.0$   &  $0.68$    &  $92.0\pm 8.0$    \\  
        $0.179$   &  $75.0\pm 4.0$   &  $0.781$   &  $105.0\pm 12.0$  \\   
        $0.199$   &  $75.0\pm 5.0$   &  $0.875$   &  $125.0\pm 17.0$  \\  
        $0.2$     &  $72.9\pm 29.6$  &  $0.88$    &  $90.0\pm 40.0$   \\   
        $0.27$    &  $77.0\pm 14.0$  &  $0.9$     &  $117.0\pm 23.0$  \\   
        $0.28$    &  $88.8\pm 36.6$  &  $1.037$   &  $154.0\pm 20.0$  \\   
        $0.352$   &  $83.0\pm 14.0$  &  $1.3$     &  $168.0\pm 17.0$  \\   
        $0.3802$  &  $83.0\pm 13.5$  &  $1.363$   &  $160.0\pm 33.6$  \\   
        $0.4$     &  $95.0\pm 17.0$  &  $1.43$    &  $177.0\pm 18.0$  \\  
        $0.4004$  &  $77.0\pm 10.2$  &  $1.53$    &  $140.0\pm 14.0$  \\    
        $0.4247$  &  $87.1\pm 11.2$  &  $1.75$    &  $202.0\pm 40.0$  \\  
        $0.4497$  &  $92.8\pm 12.9$  &  $1.965$   &  $186.5\pm 50.4$  \\  
        $0.47$    &  $89.0\pm 49.6$  &  $0.8$     & $113.1 \pm 15.2$ \\
\noalign{\smallskip}\hline
\end{tabular}
\caption{Compilation of $31$ $H(z)$ cosmic chronometers data from~\citep{Niu24}.}
\label{tab:tableA0}
\end{table}


% Bibliography

%% [A] Recommended: using JHEP.bst file
\bibliographystyle{JHEP}
\bibliography{main.bib}

\end{document}